\documentclass[aps,prl,twocolumn,groupedaddress]{revtex4}

\pdfoutput=1
\usepackage[pdftex]{graphicx}

\usepackage{latexsym,amsfonts,amssymb}
\usepackage{amsmath}

\usepackage{helvet}
\usepackage{color}

\newcommand{\be}{\begin{equation}}
\newcommand{\ee}{\end{equation}}
\newcommand{\bd}{\begin{displaymath}}
\newcommand{\ed}{\end{displaymath}}

\newcommand{\beq}{\begin{eqnarray}}
\newcommand{\eeq}{\end{eqnarray}}

\renewcommand{\phi}{\varphi}

\begin{document}

\title{Quench dynamics near a quantum critical point}

\author{C. De Grandi$^1$}
\author{V. Gritsev$^2$}
\author{A. Polkovnikov$^1$}
\affiliation{$^1$Department of Physics, Boston University, 590 Commonwealth Avenue, Boston, MA 02215, USA\\
$^2$Department of Physics, University of Fribourg, Chemin du Mus\'{e}e 3, 1700 Fribourg, Switzerland}

\begin{abstract}

We study the dynamical response of a system to a sudden change of the tuning parameter $\lambda$ starting (or ending) at the quantum critical point. In particular  we analyze the scaling of the excitation probability, number of excited quasiparticles, heat and entropy with the quench amplitude and the system size. We extend the analysis to  quenches with arbitrary power law dependence on time of the tuning parameter, showing a close connection between the scaling behavior of these quantities with the singularities of the adiabatic susceptibilities of order $m$ at the quantum critical point, where $m$ is related to the power of the quench. Precisely for sudden quenches the relevant susceptibility of the second order coincides with the fidelity susceptibility. We discuss the generalization of the scaling laws to the finite temperature quenches and show that the statistics of the low-energy excitations becomes important. We illustrate the relevance of those results for cold atoms experiments.

\end{abstract}
\maketitle

Understanding the dynamics of quantum interacting systems
 is one of the key challenge of the modern physics. The theoretical research in this area has been stimulated by the fast developments in the field of cold atom experiments~\cite{blochReview} and the emerging possibility of manipulating quantum systems~\cite{experiments1}.
In particular, the idea of suddenly change a parameter of the Hamiltonian, i.e. performing a quench, has been widely addressed within  different approaches~\cite{cardy0quenches, cardy, kollath_altman_quench, roux}.
Furthermore recently the interest in studying  quenches has been motivated  by  the questions on the  thermalization of quantum systems~\cite{thermalization}.

Quenches near quantum critical points are especially interesting because of the expected universality of the response of the system and thus the possibility of using the quench dynamics as a non-equilibrium probe of phase transitions. This universality is well established in equilibrium systems~\cite{Sachdev_book}. It has been shown that a universal scaling arises as well in the case of slow adiabatic perturbations that drive the system through a quantum critical point~\cite{toliapoint,zurekzoller}. The predicted scaling for the number of created quasiparticles with the quench rate was verified for  various specific models~\cite{dziarmaga, uma_quenchXY,sengupta2D, lviola, canevaIsing, balazs}. This analysis was also generalized to nonlinear quenches~\cite{sengupta_nonlinearPRL, optimal_passage}.

In this work we consider a $d$-dimensional system described by a Hamiltonian  $H(\lambda)=H_0+\lambda V$, where $H_0$ is the Hamiltonian corresponding to a quantum critical point (QCP) and $V$ is a relevant (or marginal) perturbation, which drives the system to a particular phase. The quench process is implemented through the parameter $\lambda$ that changes in time, according to some protocol, between the initial value $\lambda=0$ at time $t=0$, corresponding to the critical point, and the final value $\lambda_f$ at final time $t_f$~\endnote{We note that the same analysis applies to the inverse process where the quench ends at the QCP.}. We  expect two qualitatively different scenarios: (i) for fast quenches the response of the system only depends on the quench amplitude $\lambda_f$, but not on  the details of the protocol used to change $\lambda$ -  we refer to this regime as  {\em sudden} quench, (ii) for slow quenches instead the system is sensitive to the protocol but independent on $\lambda_f$, indeed the main source of non-adiabaticity are the transitions occurring at short times when the system is close to the QCP (see Ref.~\cite{nostro} for linear quenches) - we refer to this regime as  {\em slow} quench.
To differentiate between the two regimes it is convenient to introduce the transition time $t_{ tr}$ defined by the condition $d\Delta/dt\sim \Delta^2$, where $\Delta$ is a characteristic energy scale associated with the proximity to the critical point, which can be a gap or some other crossover scale~\cite{Sachdev_book}. If $t_f\ll t_{tr}$ we are dealing with a sudden quench and conversely if $t_f\gg t_{ tr}$ we are dealing with a slow quench. For linear quenches: $\lambda(t)=\upsilon t$ using that $\Delta\sim |\lambda|^{z\nu}=|\upsilon t|^{z\nu}$, we find that $t_{ tr}\sim 1/|\upsilon|^{1/(z\nu+1)}$, where $z$ and $\nu$ are the dynamical and correlation length exponent respectively~\cite{Sachdev_book}.

\textit{Sudden quenches}. For sudden changes in the parameter $\lambda$ of the Hamiltonian a natural quantity to look at is the ground state fidelity $F(\lambda,\lambda +\delta \lambda)=|\langle\Psi_0(\lambda)|\Psi_0(\lambda+\delta \lambda)\rangle|$~\cite{venuti, gu_long}, which characterizes the overlap of two ground states with slightly different values of the tuning parameter. For a quench of amplitude $\lambda_f$  (starting from $\lambda=0$) the probability of exciting the system away from the ground state is related to the fidelity simply as: $P_{\rm ex}(\lambda_f)=1-F(0,\lambda_f)^2$.
For  small quench amplitudes  we can Taylor expand $F(0,\lambda_f)^2\approx 1-|\lambda_f|^2 L^d \chi_f(0)$, where   $\chi_f(\lambda)$ is the fidelity susceptibility :
\be
\label{chif}
\chi_f(\lambda)={1\over L^d}\sum_{n\neq 0} |\langle 0|\partial_\lambda|n\rangle|^2=
{1\over L^d}\sum_{n\neq 0} { |\langle 0|V|n\rangle|^2\over (E_n(\lambda)-E_0(\lambda))^2}.
\ee
Here {$\vert n \rangle$}  is the set of eigenstates of the system evaluated at $\lambda$ and $E_n(\lambda)$ are the corresponding energies,  $L$ is the system size. From the Lehmann's representation it immediately follows that this susceptibility can be written also through the imaginary time connected correlation function $G_\lambda(\tau)=\langle V(\tau)V(0)\rangle-\langle V(0)\rangle^2$ as  $\chi_f(\lambda)={1\over L^d}\int_0^\infty d\tau\, \tau G_\lambda(\tau)$~\cite{gu_long}. Either of these representations implies that $\chi_f$ is non-negative and $L^d \chi_f$ is extensive if $V$ is a local operator and $G(\tau)$ decays faster than $1/\tau^2$ at large $\tau$. Since $G(\tau)$ is a non-negative monotonically decreasing function of $\tau$, we see that $\chi_f$ is strictly positive if $G(0)>0$, i.e. if the ground state is not the eigenstate of the operator $V$. This is expected to be true as long as $V$ does not commute with the Hamiltonian $H$ or more precisely with all integrals of motion of the system.
Near a critical point one can expect that $\chi_f$,  as any other susceptibility, has a singularity and possibly diverges. This can happen if the correlation function $G(\tau)$ falls off slower than $1/\tau^2$ at the critical point. It has been shown ~\cite{venuti,alet} that the scaling dimension of the fidelity susceptibility is $\dim[\chi_f]=d-2/\nu $. Hence for $d\nu<2$ the fidelity susceptibility diverges as: $\chi_f(\lambda)\sim |\lambda|^{d\nu-2}$ saturating at $\chi_f(0)\sim L^{2/\nu-d}$ for $|\lambda|\lesssim L^{-1/\nu}$. For $d\nu>2$ this non-analytic asymptotic with $L$ or $\lambda$ becomes subleading and the fidelity susceptibility develops a cusp singularity. Using  those scaling laws we find that  for $d\nu<2$ :
\be\label{Pex}
P_{\rm ex}\sim |\lambda_f|^2 L^{2/\nu}.
\ee
Since $P_{\rm ex}\leq 1$, this scaling must be valid only for quenches of very small amplitude $|\lambda_f|\ll L^{-1/\nu}$. This condition has a simple physical interpretation that the correlation length in the final state of the system $\xi(\lambda_f)\sim 1/|\lambda_f|^\nu$ must be big compared to the system size $\xi(\lambda_f)\gg L$. In the opposite case $\xi(\lambda_f)\lesssim L$ the system in the new ground state (with possibly a new symmetry) has a well defined order and the overlap with the old ground state is expected to be zero and consequently $P_{\rm ex}=1$.

In order to understand quenches of larger amplitude $|\lambda_f|\gg L^{-1/\nu}$, where $P_{\rm ex}$ is no longer informative, we need to extend the physical interpretation of $\chi_f$. To do this we derive the scaling (\ref{Pex}) using the adiabatic perturbation theory~\cite{toliapoint, ortiz_2008}. This theory assumes  the proximity of the system to the instantaneous ground state, but not necessarily to the initial state. Within the leading order of the perturbation theory the transition amplitude $\alpha_n(t)$ to the (instantaneous) state $|n(t)\rangle$ is given by:
\be
\alpha_n(t)\approx-\int_{0}^t dt' \langle n|\partial_{t'}|0\rangle\mathrm e^{i(\Theta_n(t')-\Theta_0(t'))},
\label{int_eq2}
\ee
where $\langle n|\partial_{t}|0\rangle =-\dot\lambda(t)\langle n| V|0\rangle/(E_n(t)-E_0(t))$ is the transition matrix element and $\Theta_n(t)=\int_0^t E_n(\tau) d\tau$ is the dynamical phase (assuming that there is no additional Berry phase). The squares of the amplitudes $|\alpha_n(t)|$ determine the transition probabilities to the excited states. Changing variables $t\to\lambda(t)$ results in the dynamical phase acquiring a prefactor $1/|\dot\lambda|$. For slow quenches, e.g. ($\lambda=\upsilon t$,  $\upsilon \to 0$) this leads to strong oscillations of the phase factor and suppression of the transitions, for sudden quenches of small amplitude this prefactor eliminates the effect of the phase so that $\alpha_n(\lambda_f)\approx -\int_0^{\lambda_f} d\lambda \langle n|\partial_{\lambda}|0\rangle$. If the matrix element is approximately independent of $\lambda$ in the interval $[0,\lambda]$, then we recover the result of the conventional first-order perturbation theory. Within this approach the probability of exciting the system is:
\be
P_{\rm ex}=\sum_{n\neq 0} |\alpha_n(\lambda_f)|^2\approx \sum_{n\neq 0}\left|\int_0^{\lambda_f} d\lambda_1
\langle 0|\partial_{\lambda_1}|n\rangle \right|^2.
\label{p_ex}
\ee
Applying the Cauchy-Schwartz inequality to the term on the right of this equation, we find $P_{\rm ex}\leq \lambda_f L^d \int_0^{\lambda_f} d\lambda' \chi_f(\lambda')$, from which, using the scaling dimension of $\chi_f$, we can recover the scaling (\ref{Pex}).

Let us look closer at the matrix elements $\langle n|\partial_\lambda|0\rangle=\langle n| V|0\rangle/(E_n-E_0)$ appearing in Eqs.~(\ref{chif}) and (\ref{p_ex}). The states $|n\rangle$ which are connected to the ground state are usually characterized by excitations of few quasiparticles, because we are typically dealing with few-particle operators. If this is the case then one can analyze the scaling of the total number of excited quasiparticles.
In integrable models, where the eigenstates of the Hamiltonian are characterized by a fixed quasiparticle number $N_n$, we can write the density of the latter as $n_{\rm ex}=1/L^d\sum_{n\neq 0} N_n |\alpha_n(\lambda_f)|^2$. For a number of models, which can be mapped to  free field theories,  $N_n=2$, i.e. only pairs of quasiparticles with opposite momenta can be excited. This is the case for the transverse field Ising model, XXZ model, Kitaev model, sine-Gordon model and  others (for a detailed analysis of the sine Gordon model see Ref.~\cite{supp}). Also for the fermionic model with two-body interactions $N_n=4$ (see e.g. Ref.~\cite{kehrein}). We expect that the universal scaling of $n_{\rm ex}$ will remain invariant even if one adds an additional small perturbation breaking the integrability of the system but keeping unchanged the universality class of the transition (for a particular bosonic model this was numerically verified in Ref.~\cite{tolianature}). The advantage of dealing with $n_{\rm ex}$ is that we do not have anymore the constraint to have vanishingly small quench amplitude $|\lambda_f|\ll L^{-1/\nu}$. Indeed exciting even a single quasiparticle creates an orthogonal state. However  physically having one quasiparticle in the system can not affect the probability to excite the next quasiparticle. Hence the validity of the many body perturbation theory should be controlled by the smallness of the intensive quantities like the density of quasiparticles. This is of course  well known in the standard linear response theory. Therefore we can safely extend the predictions of the adiabatic perturbation theory to quenches with $|\lambda_f|\gg L^{-1/\nu}$ using the scaling behavior of the fidelity susceptibility:
\beq
n_{\rm ex} \sim \left\{\begin{array}{ll} \vert \lambda_f \vert^2 L^{2/\nu-d} & \textnormal{for } \vert \lambda_f \vert\ll1/L^{1/\nu},\\
\vert \lambda_f \vert^{d \nu} & \textnormal{for } \vert \lambda_f \vert\gg 1/L^{1/\nu}.
                       \end{array}
\right.
\eeq
In Ref.~\cite{supp} we discuss in detail the relation between the scaling of $P_{\rm ex}$ and $n_{\rm ex}$ for the sine-Gordon model and show that the scaling above is indeed correct.

When well defined, $n_{\rm ex}$ is a  convenient quantity to be analyzed both theoretically and experimentally. However, if we are dealing with strongly interacting non-integrable systems which have complicated many-body spectrum the density of excitations becomes ill defined. Then one has to describe the system response by other means. Two other natural quantities, which can be defined for any Hamiltonian system are the (diagonal) entropy and the heat (or the excess energy above the new ground state). Because both quantities are extensive, it is convenient to deal with their densities:
\beq
S_d&=&-{1\over L^d} \sum_{n} |\alpha_n|^2\log|\alpha_n|^2,\\
Q&=&{1\over L^d}\sum_{n} (E_n-E_0)|\alpha_n|^2. \label{q_def}
\eeq
The scaling of $S_d$ is similar (up to possible log corrections) to that of $P_{\rm ex}$ and $n_{\rm ex}$. The advantage of the entropy over $P_{\rm ex}$ is that it is meaningful for larger amplitude quenches $|\lambda_f|\gg 1/L^{1/\nu}$, where $P_{\rm ex}\to 1$. The scaling of $Q$ is  different from the one of $n_{\rm ex}$ because of the extra energy factor in Eq.~(\ref{q_def}). This scaling is associated with the singularity of another susceptibility:
\be
\label{chie}
\chi_E={1\over L^d}\sum_{n\neq 0} { |\langle 0|V|n\rangle|^2\over E_n(\lambda)-E_0(\lambda)},
\ee
which has the scaling dimension $\dim[\chi_E]=d+z-2/\nu$. Thus for $(d+z)\nu<2$ at the QCP we have $\chi_E(0)\sim L^{2/\nu-d-z}$ and $\chi_E(\lambda)\sim |\lambda|^{(d+z)\nu-2}$ for $|\lambda|\gg L^{-1/\nu}$. As for the case of $\chi_f$, this susceptibility generically has a cusp singularity for $(d+z)\nu>2$. This scaling of $\chi_E$ implies that for $(d+z)\nu<2$ we have:
\beq\label{Qex2}
Q \sim \left\{\begin{array}{ll} |\lambda_f|^2 L^{2/\nu-d-z} & \textnormal{for } \vert \lambda_f \vert\ll1/L^{1/\nu},\\ |\lambda_f|^{(d+z)\nu}  & \textnormal{for } \vert \lambda_f \vert\gg 1/L^{1/\nu}.
                       \end{array}
 \right.
\eeq
We note that the heat is well defined for any system integrable or not and easily measurable.

\textit{Slow quenches.} As we already argued in this case it is important to know the asymptotic time dependence of the quench parameter $\lambda(t)$ near the critical point. We will focus on the power law asymptotic:
\be\label{Gen_quench}
\lambda(t)=\upsilon {t^r\over r!}\Theta(t),\;\; \textnormal{or}\;\; \lambda(t)=\upsilon {(t_f-t)^r\over r!}\Theta(t_f-t),
\ee
where $\upsilon$ is a small parameter and $\Theta$ is the step function. For $r=0$ the parameter $\upsilon$ plays the role of the quench amplitude $\lambda_f$, for linear quenches  $r=1$  it plays the role of the quench rate, for quadratic quenches $r=2$ it is the acceleration and so on. In all cases the limit $\upsilon\to 0$ is the adiabatic one and thus $\upsilon$ plays the role of the adiabaticity parameter.
Using once again the adiabatic perturbation theory (see Eq.~(19) in Ref.~\cite{kolkata} or Eq.~(18) in Ref.~\cite{ortiz_2008}) we find that the transition amplitude to the instantaneous excited state $|n\rangle$
is dominated by the lowest non-vanishing time derivative of $\lambda(t)$ at the critical point:
\be
|\alpha_n|^2\approx \upsilon^2 {|\langle n|\partial_\lambda|0\rangle|^2\over (E_n(0)-E_0(0))^{2r}}=
\upsilon^2 {|\langle n|V|0\rangle|^2\over (E_n(0)-E_0(0))^{2r+2}}.
\ee
Therefore perturbatively we find:
\be
P_{\rm ex}= \sum_{n\neq 0}|\alpha_n|^2\approx \upsilon^2 L^d \chi_{2r+2}(0),
\ee
where we introduced the generalized adiabatic susceptibility of  order $m$:
\be
\chi_{m}(\lambda)=\frac{1}{L^d}\sum_{n\neq 0}{|\langle n| V|0\rangle|^2\over (E_n(\lambda)-E_0(\lambda))^m}.
\ee
Clearly $\chi_1=\chi_E$, $\chi_2=\chi_f$, $\chi_4$ is the susceptibility which describes the scaling of $P_{\rm ex}$ and $n_{\rm ex}$ for slow linear quenches and so on. All these susceptibilities are also related to the connected correlation function: $\chi_m(\lambda)=1/[L^d(m-1)!]\int_0^\infty \tau^{m-1} G_\lambda(\tau) d\tau$. It is easy to see that $\dim[\chi_m]=d-2/\nu-z(m-2)$ which implies that for small $\upsilon$ and $d\nu\leq 2(1+z\nu r)$ we have:
\be
P_{\rm ex}\approx \upsilon^2 L^{2/\nu+2 z r}.
\ee
This scaling is expected to be valid for $|\upsilon|\ll 1/L^{1/\nu+zr}$. For larger $\upsilon$ as in the case of sudden quenches one can consider the scaling of other quantities like $n_{\rm ex}$:
\beq
n_{\rm ex} \sim \left\{\begin{array}{ll} \upsilon^2 L^{2/\nu-d+2 z r} & \textnormal{for } |\upsilon|\ll 1/L^{1/\nu+zr} ,\\
|\upsilon|^{d\nu/(z\nu r+1)}  & \textnormal{for } |\upsilon|\gg 1/L^{1/\nu+zr}.
                       \end{array}
 \right.
\eeq
Similar scaling laws are valid for the entropy. We point out that, for $r=1$, this scaling for $n_{\rm ex}$ agrees with the one predicted earlier in Refs.~\cite{toliapoint, zurekzoller}, and for nonlinear quenches, $r>1$, it agrees  with the results of Refs.~\cite{krishnendu, optimal_passage}. If the quench ends at the critical point then one finds that the scaling of the heat is described by $\chi_{2r+1}$: $Q\sim \upsilon^2 \chi_{2r+1}$ so that:
\beq
Q \sim \left\{\begin{array}{ll} \upsilon^2 L^{2/\nu-d+z (2r-1)} & \textnormal{for } |\upsilon|\ll 1/L^{1/\nu+zr} ,\\|\upsilon|^{(d+z)\nu/(z\nu r+1)} & \textnormal{for } |\upsilon|\gg 1/L^{1/\nu+zr}.
                       \end{array}
 \right.
\eeq
We note that these scaling results can be applied to generic gapless systems in the absence of a quantum critical point by formally sending $\nu\to\infty$.

So far we focused only on quenches at zero temperatures, where the system is initially prepared in the ground state. However, it is well known that in equilibrium the properties of the system are governed by the QCP well into the finite temperature domain, so called quantum critical region~\cite{Sachdev_book}. We can expect that the same is true for the dynamics. Motivated by the cold atom experiments we consider an isolated system but initially prepared at some finite temperature $T$,  similarly to  Refs.~\cite{tolianature, cardy} (alternative scenarios involving a coupling to a thermal bath were discussed in Ref.~\cite{Patane_short}). In Ref.~\cite{supp} we analyze  the case of thermal quenches for systems of free massive bosons and fermions, we refer there for a detailed derivation. Here we simply mention the results, which for these limits are simple and intuitive. Thus if the excitations are bosonic (fermionic) in nature then the number of  quasiparticles excited into the mode with momentum $q$ during the quench, $n_{\rm ex}(q)$, at finite temperature is related to the one at zero temperature, $n_{\rm ex}^0(q)$, via: $n_{\rm ex}(q)=n_{\rm ex}^0\coth{(\epsilon_q^0/2T)}$ ($n_{\rm ex}(q)=n_{\rm ex}^0\tanh{(\epsilon_q^0/2T)}$), where $\epsilon_q^0$ is the initial energy of the quasiparticle. This result is valid for an {\em arbitrary} time dependence of the tuning parameter (see \cite{tolianature} for the bosonic case). At large temperatures $T\gg \epsilon_q^0$, the hyperbolic cotangent (tangent) factor gives  enhancement (suppression) of the density of excited quasiparticles, so that in the resulting scaling relations one has to substitute $d\to d-z$ ($d\to d+z$) and simultaneously multiply (divide) by the temperature, e.g. $n_{\rm ex}\sim T^{\pm 1} |\upsilon|^{(d\mp z)\nu/(z\nu r+1)}$, where the upper (lower) sign corresponds to bosonic (fermionic) systems (the same is true for the scaling of $Q$). Interestingly the same hyperbolic cotangent (tangent) factors appear in usual fluctuation-dissipation relations. Their origin simply reflects the bosonic (fermionic) statistics and is not tied to the assumption of being in the linear response regime. As in the zero temperature case one expects these leading scaling asymptotics to be valid as long as the corresponding exponents are less than two, otherwise they become subleading.  An interesting question is what happens at finite temperatures if the low-energy excitations have fractional statistics or no well defined statistics at all. Additional ingredients from the low-energy theory should enter the scaling relations compared to the zero temperature case where the statistics is unimportant. At the moment this question remains open.

Finally let us comment that the predictions of this work can be directly probed experimentally. In the context of cold atom systems one can perform a sudden quench starting from the critical point. E.g. to probe the sine-Gordon model one can suddenly turn on a commensurate optical lattice potential in a 1D system of interacting bosons \cite{nostro}. Then one can adiabatically drive the system to the regime where the  excitations are easily measurable, e.g. increasing the amplitude of the lattice potential to a large value where the excitations simply correspond to sites with double or zero occupancy. Similarly one can measure the heat by performing e.g. a cyclic process and measuring the excess energy with time of flight experiments.

In conclusion, we derived the scaling relations for different observables in a system quenched away from a QCP. We found that the scaling laws of the heat, entropy, probability of excitations are universal and determined by the critical exponents $z$ and $\nu$ of the system. We discussed a close connection between sudden and slow quenches and showed that the  universal scaling of various quantities can be understood through the singularities of generalized adiabatic susceptibilities at the QCP. We also consider the case of finite temperature quenches and argued that the statistics of the quasiparticles strongly affects the scaling results. We believe those predictions can be directly probed in cold atoms systems.

We acknowledge useful discussions with   D. Baeriswyl, R. Barankov,  B. Gut, G. Roux, K.~Sengupta, and A.~Silva. The work was supported by  AFOSR YIP, NSF: DMR 0907039, and Sloan Foundation. C.~D.~G. acknowledges the support of I2CAM: DMR-0645461. V.G. was supported by Swiss NSF.

\bibliography{quench}

\end{document}